\documentclass[conference]{IEEEtran}
\IEEEoverridecommandlockouts
\usepackage{cite}
\usepackage{amsmath,amssymb,amsfonts}
\usepackage{algorithmic}
\usepackage{graphicx}
\usepackage{textcomp}
\usepackage{xcolor}
\usepackage{float}
\usepackage[caption = false]{subfig}
\usepackage{siunitx}
\usepackage{scalerel}
\usepackage{multirow}
\usepackage{pgfplots}
\usepackage{soul}
\usepackage{acronym}
\pgfplotsset{compat=newest}
\usepackage[center]{caption}
\usepackage{eurosym}
\usepackage{multicol}
\usepackage{booktabs}
\usepackage{svg}
\usepackage{enumitem,kantlipsum}
\usepackage{adjustbox}

\DeclareSIUnit{\var}{Var}
\DeclareSIUnit{\va}{VA}
\DeclareSIUnit{\pu}{p.u.}
\DeclareSIUnit{\deg}{deg}
\DeclareSIUnit{\MW}{MW}
\DeclareSIUnit{\MWh}{MWh}
\DeclareSIUnit{\min}{min}
\DeclareSIUnit{\t}{t}
\DeclareSIUnit{\EUR}{\text{\euro}}
\DeclareSIUnit{\kn}{knts}

\acrodef{bess}[BESS]{Battery Energy Storage System}
\acrodef{eedi}[EEDI]{Energy Efficiency Design Index}
\acrodef{imo}[IMO]{International Maritime Organization}
\acrodef{epla}[EPLA]{Electric Power Load Analysis}
\acrodef{dg}[DG]{Diesel Generator}
\acrodef{seemp}[SEEMP]{Ship Energy Efficiency Management Plan}
\acrodef{uc}[UC]{Unit Commitment}
\acrodef{milp}[MILP]{Mixed Integer Linear Programming}
\acrodef{ghg}[GHG]{Global Greenhouse Gases}
\acrodef{nox}[$NO_X$]{Nitrogen Oxide}
\acrodef{sox}[$SO_X$]{Sulphur Oxide}
\acrodef{pm}[PM]{Particulate Matter}
\acrodef{un}[UN]{United Nation}
\acrodef{vlsfo}[VLSFO]{Very Low Sulphur Fuel Oil}
\acrodef{egcs}[EGCS]{Exhaust Gas Cleaning Systems}
\acrodef{eedi}[EEDI]{Energy Efficiency Design Index}
\acrodef{eexi}[EEXI]{ Energy Efficiency Existing Ship Index}
\acrodef{cii}[CII]{Carbon Intensity Indicator}
\acrodef{gt}[GT]{Gross Tonnage}
\acrodef{mc}[MC]{Markov Chain}
\acrodef{sfoc}[SFOC]{Specific Fuel Oil Consumption}
\acrodef{soc}[SOC]{State Of Charge}
\acrodef{sog}[SOG]{Speed Over Ground}
\acrodef{oc}[OC]{Operating Condition}
\acrodef{igsc}[IGSC]{Instantaneous Growing Stream Clustering}
\acrodef{swbs}[SWBS]{Ship Work Breakdown Structure}
\acrodef{sps}[SPS]{Shipboard Power System}
\acrodef{mepc}[MEPC]{Marine Environment Protection Committee}
\acrodef{dcs}[DCS]{Data Collection System}
\acrodef{eeoi}[EEOI]{Energy Efficiency Operational Indicator}
\acrodef{solas}[SOLAS]{Safety Of Life At Sea}
\acrodef{sgp}[SGP]{State Growth Parameter}
\acrodef{eca}[ECA]{Emission Control Area}
\acrodef{aes}[AES]{All Electric Ship}
\acrodef{pms}[PMS]{Power Management System}
\acrodef{iacs}[IACS]{International Association of Classification Societies}
\acrodef{pemfc}[PEMFC]{Proton Exchange Membrane Fuel Cell}
\acrodef{fc}[FC]{Fuel cell}
\acrodef{dod}[DOD]{Depth Of Discharge}
\acrodef{loh}[LoH]{Level of Hydrogen}
\acrodef{ccus}[CCUS]{Carbon Capture, Usage and Storage}
\acrodef{ffr}[FFR]{Fuel Flow Rate}
\acrodef{hfr}[HFR]{Hydrogen Flow Rate}
\acrodef{etd}[ETD]{Energy Taxation Directive}
\acrodef{ets}[ETS]{Emissions Trading System}
\acrodef{rl}[RL]{Reinforcement Learning}
\acrodef{mpc}[MPC]{Model Predictive Control}

\begin{document}

\title{Performance Investigation of an Optimal Control Strategy for Zero-Emission Operations of Shipboard Microgrids}

\author{\IEEEauthorblockN{Fabio D'Agostino, Marco Gallo, Matteo Saviozzi, Federico Silvestro}
\IEEEauthorblockA{\textit{Electrical, Electronics and Telecommunication Engineering and Naval Architecture Department - DITEN}\\
    \textit{Universita degli Studi di Genova} Genova, Italy\\
    fabio.dagostino@unige.it, marco.gallo@edu.unige.it, matteo.saviozzi@unige.it, federico.silvestro@unige.it}
}
\maketitle

\begin{abstract}
This work introduces an efficient power management approach for shipboard microgrids that integrates diesel generators, a fuel cell, and battery energy storage system. 
This strategy addresses both unit commitment and power dispatch, considering the zero-emission capability of the ship, as well as optimizing the ship's speed. The optimization is done through mixed integer linear programming with the objective of minimizing the operational cost of all the power resources. Evaluations are conducted on a notional all-electric ship, with electrical load simulated using a Markov chain based on actual measurement data. The findings underscore the effectiveness of the proposed strategy in optimizing fuel consumption while ensuring protection against blackout occurrences.
\end{abstract}


\section{Introduction} \label{Intro}
The reduction of CO\textsubscript{2} emissions from ships is one of the main topics concerning the maritime sector. Starting from January 1\textsuperscript{st} 2023, all ships are required to calculate their \ac{eexi} to assess energy efficiency. This measure also involves collecting data for reporting their annual operational \ac{cii} and \ac{cii} ratings \cite{IMOCII}.

This aligns with the European Union’s Fit for 55 climate package of legislative proposals \cite{fit455}, which includes initiatives aimed at reducing \ac{ghg} emissions by 55\% by 2030 compared to 1990 levels.

Several recent papers suggest employing \ac{bess} to improve efficiency in \ac{aes} powered by \acp{dg}. In \cite{BESSuse}, various functions for \ac{bess}  are described, with strategic loading being an interesting function that optimizes the operating point of the \acp{dg}.

Starting from January 2026, the Norwegian Maritime Authority will require that all navigation in the fjord to be zero-emission \cite{ZeroEmissionNor}. 

Efficiency and zero-emission capability must be combined with a robust and secure management of the generation system.
According to the \ac{iacs} guidelines, in the event of a failure of one generating unit, the system must be able to avoid the blackout \cite{IACS}. Hence, the security constraints must account for the primary limitations of each generating unit, including both the maximum overload capacity and the maximum allowable load step that a generator can handle during emergencies.

In \ac{aes} one of the main challenges lies in designing an effective \ac{pms} strategy that coordinates the power sources to achieve efficient and robust operation.

In the existing literature, several authors have proposed different power management strategies.
In \cite{ScheduledEMS}, the scheduling methodology outlined optimizes ship operations with the goal of minimizing fuel consumption and diesel generator emissions.
In \cite{RLandMPC},  two supervisory optimization-based power and energy management control systems have been proposed with \ac{mpc} and \ac{rl} approaches.

Furthermore, the use of fuel cells to ensure zero-emission capability has been suggested by various authors.
In \cite{ZeroEmissionFC}, a zero-emission hybrid ship based on \ac{fc}, batteries and cold ironing is proposed.
In \cite{RealTimeOpt}, a sizing method for the energy storage system in a \ac{fc} hybrid ferry is proposed, along with a real-time optimization control strategy.

In this paper, an optimal power management strategy for shipboard microgrids equipped with \acp{dg}, a fuel cell and a \ac{bess} is proposed. 
The algorithm, based on an optimization \ac{milp} problem, provides both the \ac{uc} and the economic dispatch of all generating units.

A preliminary version of this approach has been presented in \cite{ESARS}. The innovative contributions of this work are represented by: the modeling of \ac{pemfc} in the \ac{sps} and the inclusion of the generation cost of all power sources in the objective function.

The rest of the paper is organized as follows: Section II introduces the System Modelling adopted for the \ac{pms}, Section III provides the Optimization Problem, Section IV reports Simulation and Results Analysis, while the Conclusions are reported in Section V.

\section{System Modelling}
\begin{figure}[h]
    \centering
        \includegraphics[width=\columnwidth]{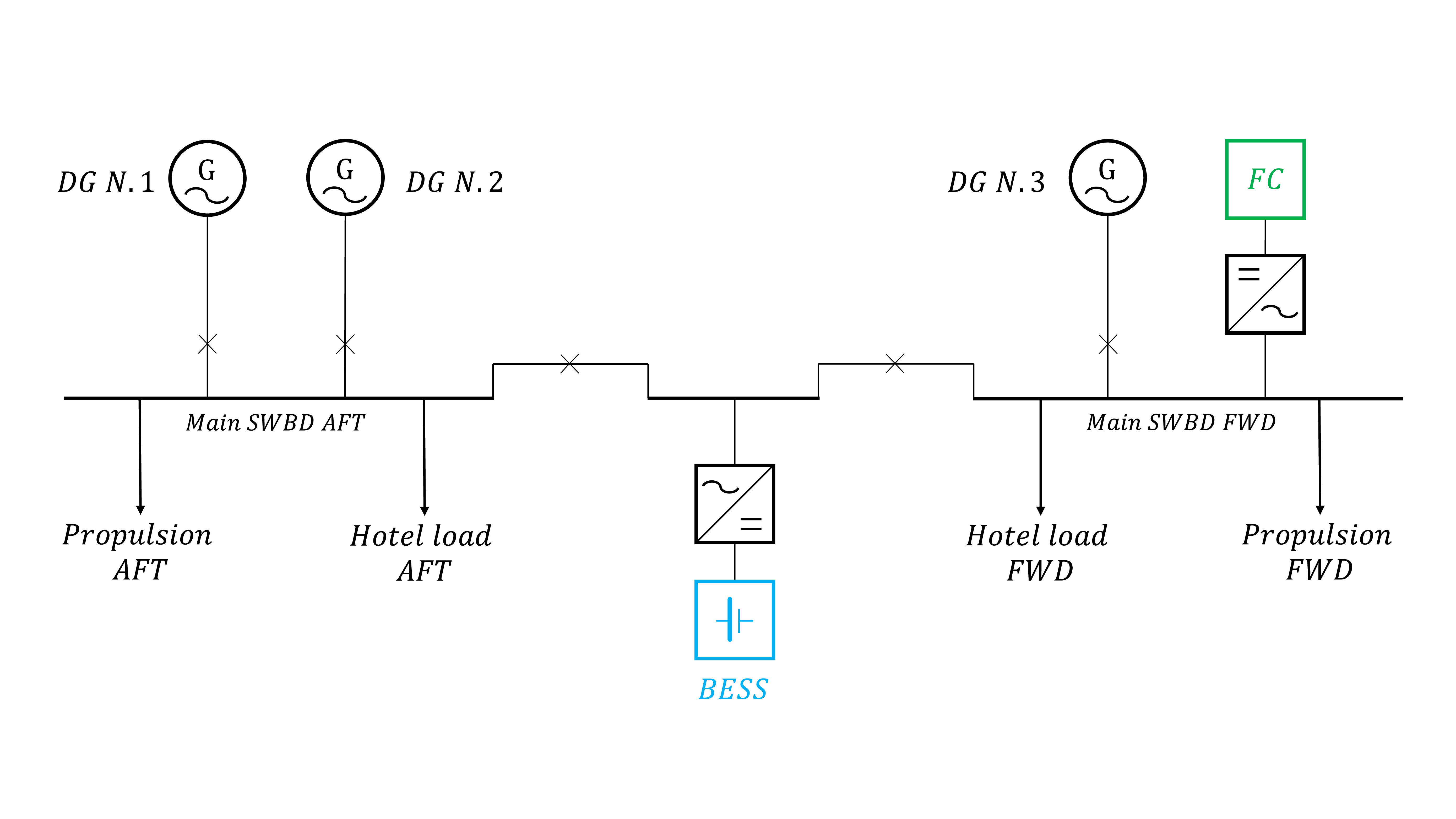}
    \caption{System Architecture, adapted from \cite{ESARS}}\label{fig:SystemArchitecture}
\end{figure}

Figure~\ref{fig:SystemArchitecture} reports the notional architecture of the selected cruise ship. The generating resources of the ship are composed of \acp{dg}, a \ac{pemfc} and a \ac{bess} that are based on real components.

In an \ac{aes} the primary load is typically the propulsion system, which correlates with the ship's speed. Additional propulsive loads such as HVAC, galley, and accommodation are encompassed within the hotel load \cite{EPLAiees}.
These loads are assessed in the \ac{epla} of the ship, categorized based on the \acp{oc} of the vessel \cite{Doerry}.

\subsection{\ac{dg} Fuel Consumption and \ac{bess}} \label{DGmodel}
The \ac{dg} fuel consumption is modelled by a linearized \ac{sfoc} curve and the \ac{soc} of the battery is modelled according to the following equation ($\forall \; t = 1:T$), as in \cite{ESARS}: 
\begin{equation}
    SOC(t+1) = SOC(t) + \biggl( P_{b}^{c}(t) \eta_{c} - 
    \begin{aligned}[t]
    & \left. \frac{P_{b}^{d}(t)}{\eta_{d}}  \right) \frac{\Delta t}{E_n} \label{eq:SOCbatteria}
    \end{aligned}
\end{equation}
where $E_n$ [\SI{}{\kWh}] is the rated energy of the battery, $P_{b}^{c}(t)$ [\SI{}{\kW}] and $P_{b}^{d}(t)$ [\SI{}{\kW}] are respectively the charge and discharge power variables of the battery at time $t$, $\Delta t$ is the granularity of the control and $T$ is the horizon of the optimization (notice that $t = 1:T$ stands for $t = 1,...,T$).

\subsection{\ac{pemfc} and Hydrogen Storage}
\begin{figure}[h]
	\centering
        \includegraphics[width=\columnwidth]{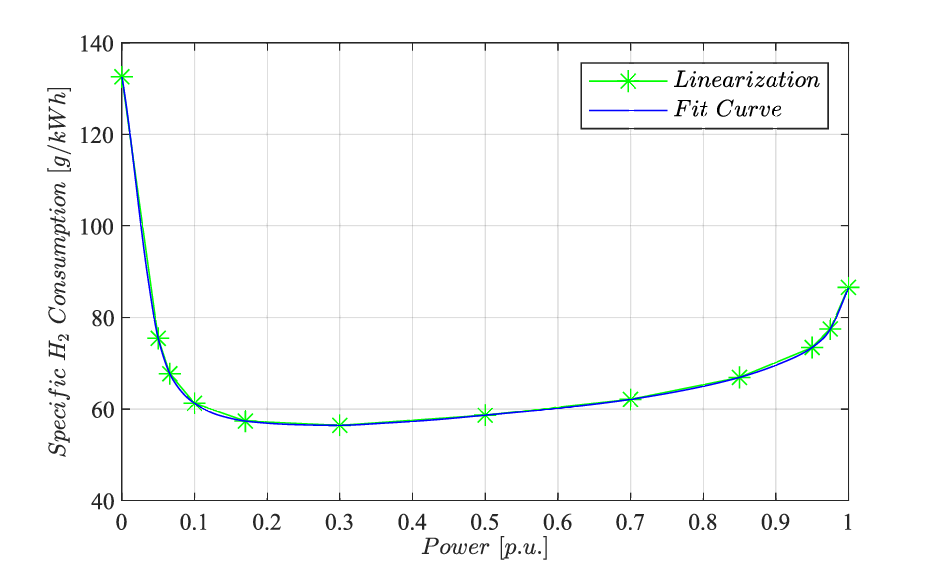}
	\caption{Specific hydrogen consumption curve.}\label{fig:SHCcurve}
\end{figure}

The specific hydrogen consumption is modeled by a linearized specific hydrogen consumption curve, as reported in Fig. \ref{fig:SHCcurve}. This curve is derived from \cite{FClin} and subsequently linearized using the methodology previously described for diesel generators.

Linearization points are selected by identifying the regions with the most significant derivative variations to minimize approximation errors.

The \ac{loh} is modelled according to the following equation:
\begin{equation}
    \begin{aligned}[t]
    LoH(t+1) = LoH(t) - \frac{\dot m_{H_2}(t)  \Delta t}{M_{H_2}} \quad \forall \; t = 1:T\label{eq:SOCbatteria}
    \end{aligned}
\end{equation}
where $M_{H_2}$ [\SI{}{\tonne}] is the total mass of H\textsubscript{2} available and $\dot m_{H_2}$ [\SI{}{t/h}] is the H\textsubscript{2} flow rate variable. The latter is derived from the specific hydrogen consumption curve shown in Fig. \ref{fig:SHCcurve}.

\section{Optimization Problem}
The proposed methodology is based on a \ac{milp} optimization algorithm implemented in MATLAB/General Algebraic Modelling System (GAMS) environment.
Several constraints are implemented in the algorithm to model the \acp{dg}, the \ac{bess}, the \ac{pemfc} and to ensure the system's security.

The objective function is formulated as:
\begin{equation}
    \begin{aligned}[t]
    &  \min \sum\limits_{t=1}^{T} \left( \sum\limits_{i=1}^{N} \dot m_{f,i}(t) \Bigl(c_f + e_f c_{CO_2}\Bigl) + c_{H_2} \dot m_{H_2}(t) + \right.\\
    &  + c_i  u_i(t) + c_b  \Bigl(1-SOC(t)\Bigl) \Biggl) \Delta t \label{eq:ObjFcn}\\
    \end{aligned}
\end{equation}
where $\dot m_{f,i}(t)$ is the fuel flow rate of $i$-th \ac{dg}, $c_{f}$ is the cost of the fuel [\SI[per-mode=symbol]{}{\EUR\per\kg}], $e_f$ is the emission factor of CO\textsubscript{2}, $c_{CO_2}$  is the cost of CO\textsubscript{2} [\SI[per-mode=symbol]{}{\EUR\per\kg}], $c_{H_2}$ is the cost of the hydrogen [\SI[per-mode=symbol]{}{\EUR\per\kg}], $c_i$ represents the start-up cost of the $i$-th generator and $c_b$ is the cost associated with the \ac{dod} of the battery ($1-SOC(t)$).
The optimization algorithm minimizes the total cost [\SI{}{\EUR}], which is composed of three parts: the cost of fuel and hydrogen, the start-up cost, and the cost related to battery degradation. The latter accounts for \ac{dod} and, therefore, considers battery degradation aspects. The term $c_b$ assigns a cost to battery management.

The cost function \eqref{eq:ObjFcn} is subjected to several constraints. In \cite{ESARS}, it is reported a detailed description of each constraint.

\section{Simulations and Results Analysis} \label{SimAndRes}
The proposed algorithm has been validated through the simulation of a shipboard microgrid consisting of three \acp{dg}, a \ac{pemfc} and a \ac{bess} (see Fig.\ref{fig:SystemArchitecture}).
The nominal power of each generating source is based on a previous work \cite{IPD}.

Since the ship electrical load profile was not available, it was modelled by simulating a generic operating profile, as in \cite{ESARS}.

\begin{figure}[h]
	\centering
        \includegraphics[width=8.5cm]{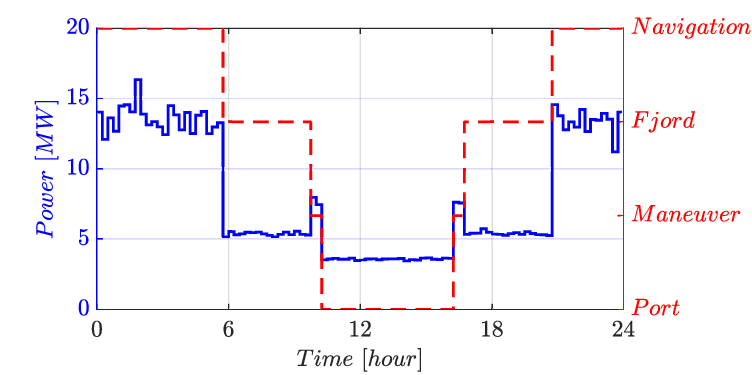}
	\caption{\ac{oc} and power profile.}\label{fig:Power_OCplot}
\end{figure}

Figure~\ref{fig:Power_OCplot} shows the load profile obtained through the simulation. In this case, the granularity of the load modelling is equal to 15 min and T = 24 hour. 
The goal is to replicate the typical navigation of the cruise ship within a Norwegian fjord. The maximum speed variation is \SI{2}{\kn} in the Fjord \ac{oc}.

The simulation parameters are listed in Table~\ref{tab:SimParam}.

\begin{table}[h]
		\caption{Parameters of the simulations.}
		\label{tab:SimParam}
		\renewcommand{\arraystretch}{1.4}
        \begin{tabular}{ccl}
        \hline \hline
        Parameter                & Value                                        & Description\\
        \hline
        $P_{DG,i = 1}^n$         & \SI{5.04}{\MW}                               & Rated power of i = 1 \ac{dg}\\
        $P_{DG,i = 2,3}^n$       & \SI{6.72}{\MW}                               & Rated power of i = 2,3 \ac{dg}\\
        $t_{min_u,i}$            & \SI{30}{\min}                                & Min up-time of i-th gen unit\\
        $t_{min_d,i}$            & \SI{30}{\min}                                & Min down-time of i-th gen unit\\
        $c_{g,min}$              & 0                                            & Min power of gen unit\\
        $c_{g,max}$              & 1                                            & Max power of gen unit\\
        $\Delta P_{r_u,i}$       & \SI[per-mode=symbol]{10}{\MW\per\min}        & Max gen unit power ramp-up limit\\
        $\Delta P_{r_d,i}$       & \SI[per-mode=symbol]{10}{\MW\per\min}        & Max gen unit power ramp-down limit\\
        $\alpha_{g}$             & 1.1                                          & Max gen unit overload in emergency\\
        $\beta_{g}$              & 0.33                                         & Max gen unit step in emergency\\
        $c_i$                    & \SI{200}{\EUR}                               & gen unit start-up cost\\
        $c_f$                    & \SI[per-mode=symbol]{0.861}{\EUR\per\kg}     & fuel cost\\  
        $c_{CO_2}$               & \SI[per-mode=symbol]{0.3}{\EUR\per\kg}       & CO\textsubscript{2} cost\\
        $e_f$                    & 3.206                                        & Emission factor CO\textsubscript{2}\\
        $P_{FC}^n$               & \SI{6}{\MW}                                  & Rated power of the \ac{pemfc}\\
        $M_{H_2}$                & \SI{10}{\tonne}                              & Mass of H\textsubscript{2} in the storage\\
        $c_{H_2}$                & \SI[per-mode=symbol]{5.176}{\EUR\per\kg}     & H\textsubscript{2} cost\\
        $n_i^{DG}$               & 10                                           & Number of \ac{sfoc} intervals curve\\
        $n_i^{FC}$               & 11                                           & Number of specific H\textsubscript{2} intervals curve\\
        \hline
        $P_{b}^{n}$              & \SI{5.35}{\MW}                                  & \ac{bess} nominal power\\
        $E_{n}$                  & \SI{5.35}{\MWh}                                 & \ac{bess} nominal energy\\
        $SOC(1)$                 & 50\%                                         & Initial \ac{soc} of the battery\\
        $SOC(T)$                 & 50\%                                         & Final \ac{soc} of the battery\\
        $SOC_{max}$              & 80\%                                         & Max \ac{soc} of the battery\\
        $SOC_{min}$              & 20\%                                         & Min \ac{soc} of the battery\\
        $\eta_d$                 & 92\%                                         & \ac{bess} discharge efficiency\\
        $\eta_c$                 & 95\%                                         & \ac{bess} charge efficiency\\
        $c_{c,min}$              & 0                                            & Min charging C-rate\\
        $c_{c,max}$              & 1                                            & Max charging C-rate\\
        $c_{d,min}$              & 0                                            & Min discharging C-rate\\
        $c_{d,max}$              & 2                                            & Max discharging C-rate\\
        $\alpha_{b}$             & 3                                            & Max \ac{bess} overload in emergency\\
        $c_b$                    & \SI[]{5}{\EUR}                               & Battery utilization cost\\
        \hline
        $M$                      & $10^9$                                       & Big M parameter\\
        $N$                      & 5                                            & Number of gen unit\\
        $D$                      & 197.9                                        & Total distance travelled by the ship\\
        $CII_{max}$              & 13                                           & Maximum allowable \ac{cii}\\
        \hline \hline
        \end{tabular}
\end{table}

\subsection{Sensitivity Analysis}
To test the performance of the algorithm, a sensitivity analysis is performed. Different sizes of \ac{bess} and \ac{fc} have been exploited. The range goes from 0 to 10 \SI{}{\MW} for both generating units. The aim is to evaluate the zero emission capability in different configuration of the \ac{sps}.

\begin{figure}[h]
	\centering
        \includegraphics[width=\columnwidth]{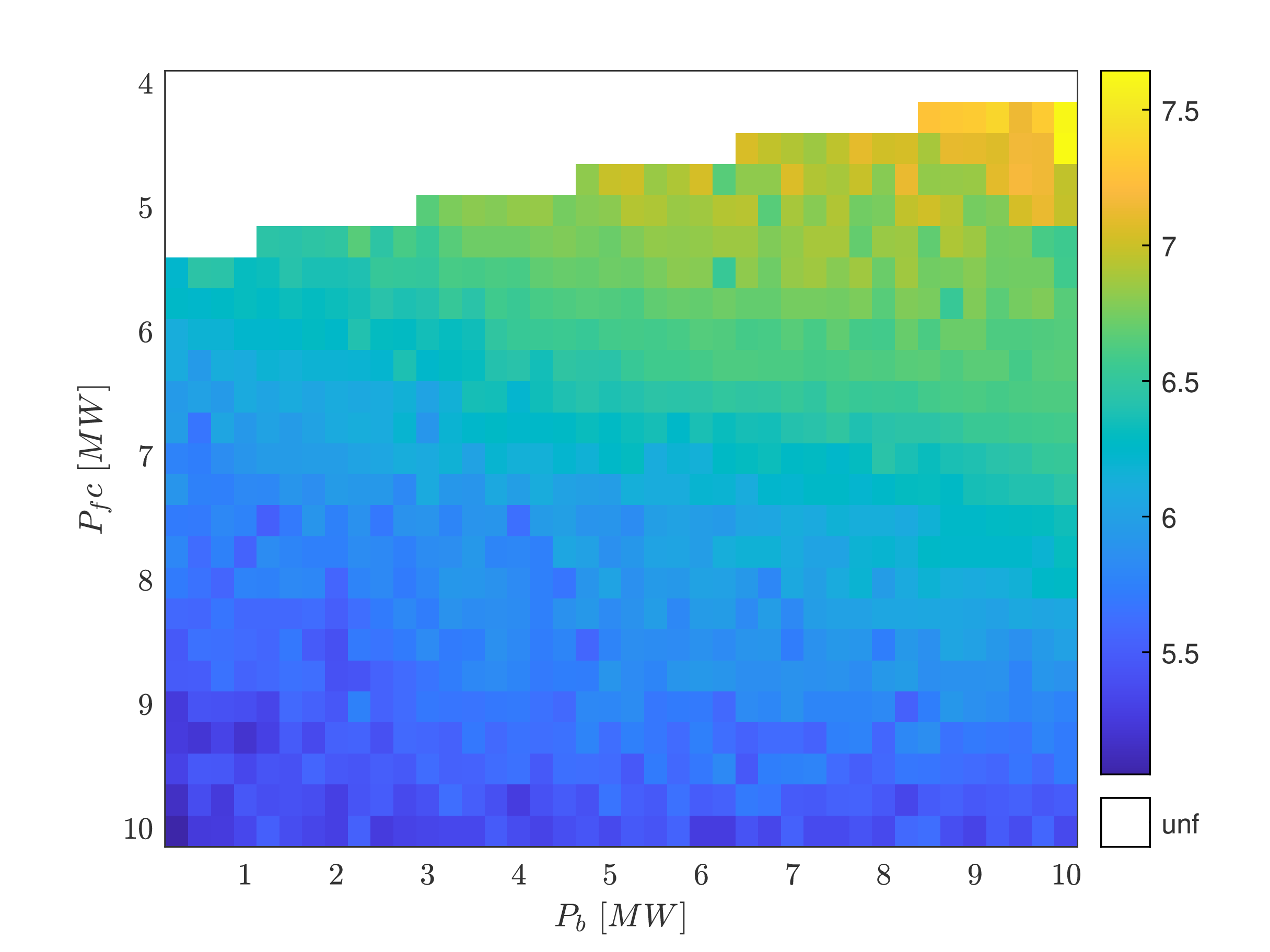}
	\caption{\ac{cii} comparison.} \label{fig:CIIsensitivity}
\end{figure}

Figure \ref{fig:CIIsensitivity} presents the results in terms of \ac{cii} rating for all the tested configurations.
The white squares are referred to an unfeasible solution of the optimization.
The lowest \ac{cii} ratings are associated with high power ratings of the \ac{fc} and battery.

\begin{figure}[h]
	\centering
        \includegraphics[width=\columnwidth]{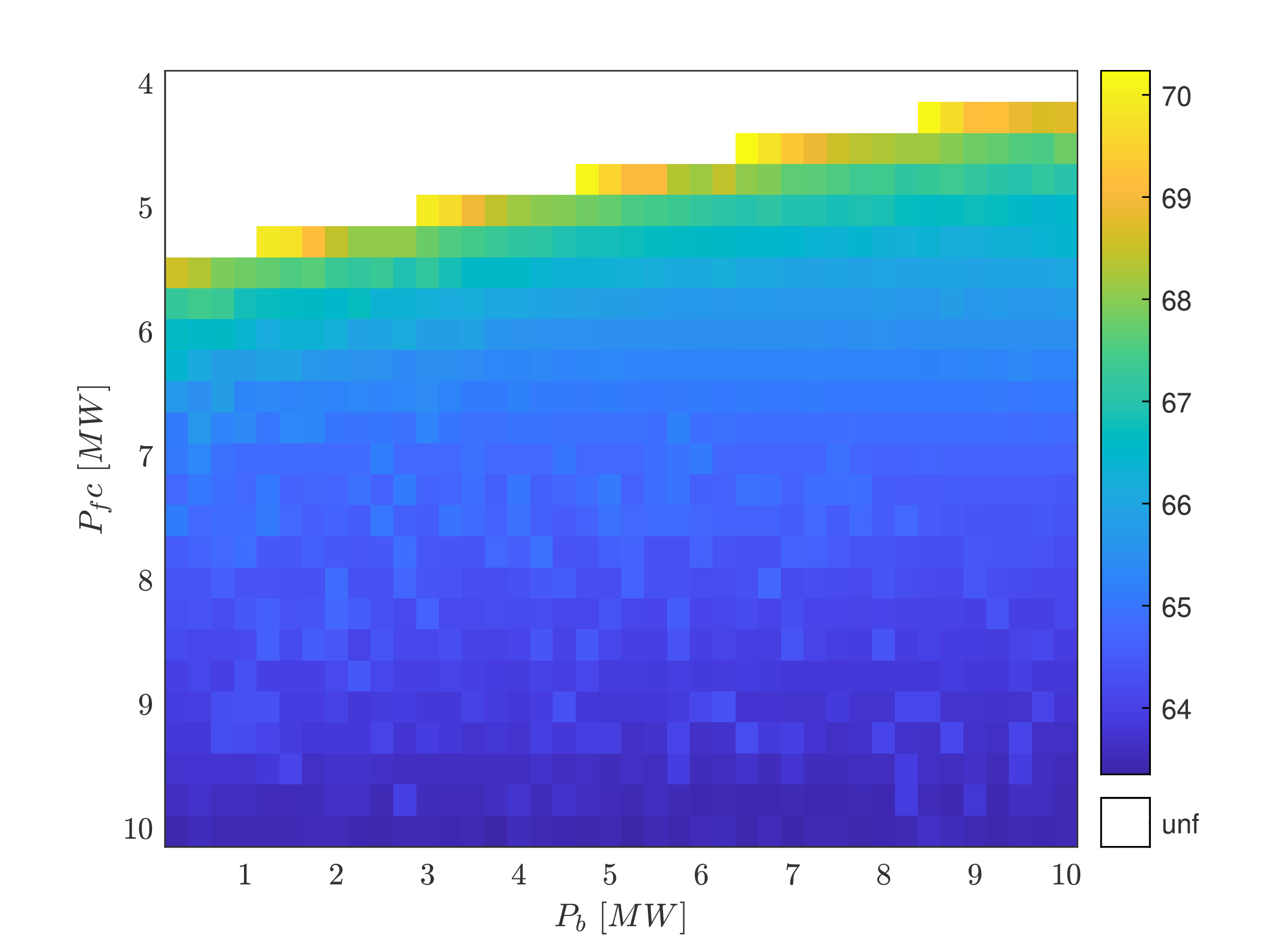}
	\caption{Cost comparison.} \label{fig:COSTsensitivity}
\end{figure}

In terms of total operational cost, the Fig. \ref{fig:COSTsensitivity} reports that the lowest values are associated with high power ratings of the \ac{fc} and battery.

\begin{figure}[h]
	\centering
        \includegraphics[width=\columnwidth]{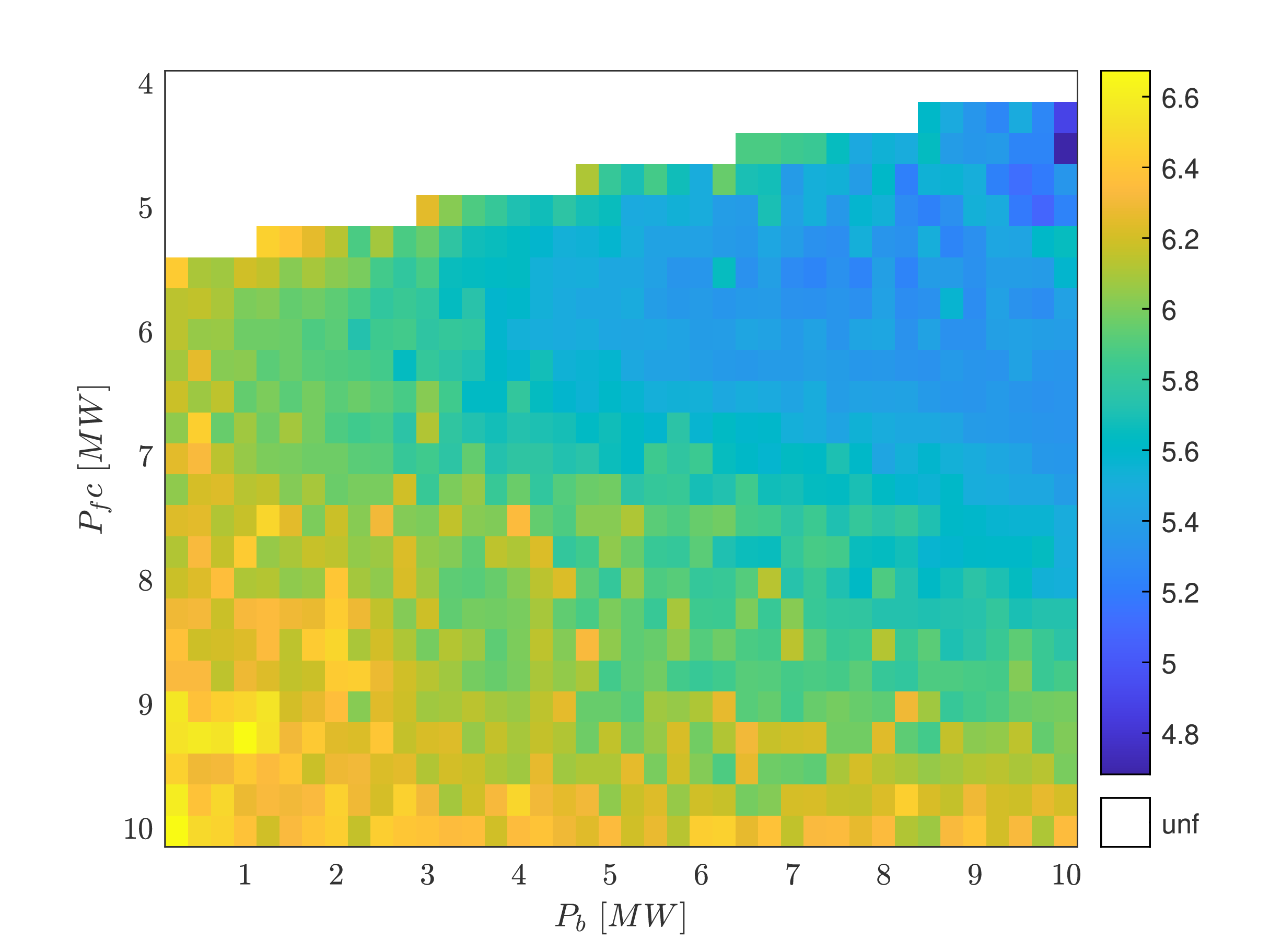}
	\caption{Hydrogen consumption comparison.} \label{fig:FuelFCsensitivity}
\end{figure}

A different pattern can be seen for the hydrogen consumption in Fig. \ref{fig:FuelFCsensitivity}. 
In this case, lower value of hydrogen consumption are given for low \ac{fc} power rating and high \ac{bess} energy rating.

\begin{figure}[h]
	\centering
        \includegraphics[width=\columnwidth]{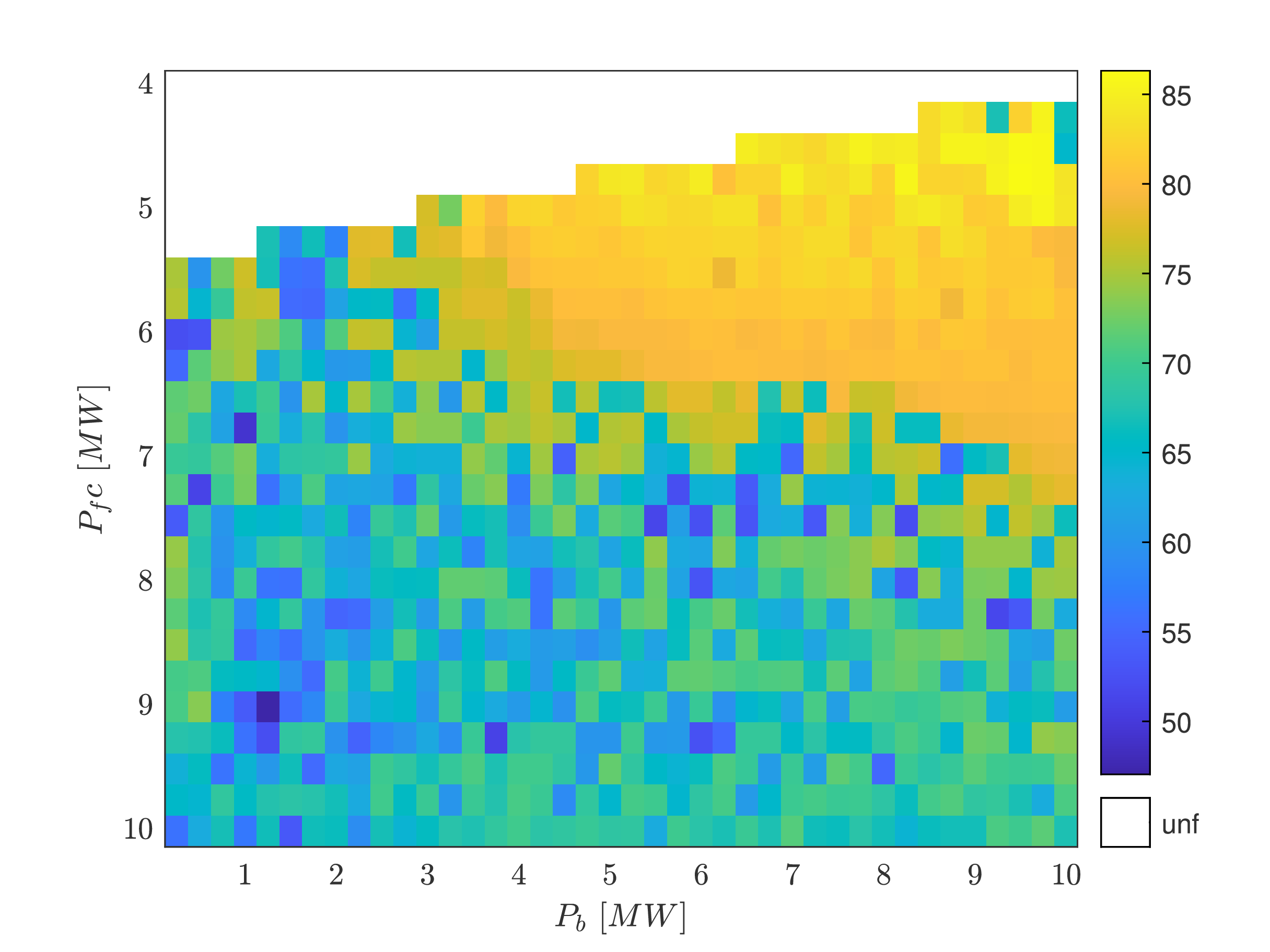}
	\caption{\ac{dg} average load factor comparison.} \label{fig:LFsensitivity}
\end{figure}

Finally, a comparison in terms of average load factor of the \acp{dg} is computed. Figure \ref{fig:LFsensitivity} shows that value next to the lowest \ac{sfoc} are given for approximately $P_{fc} = 5.75 \; MW$ and $E_b = 4.75 \; MWh$.

It is worth noting that the lowest \ac{fc} power rating where the optimization problem provides a feasible solution is $P_{fc} = 4.25 \; MW$. While for the battery, there are no issue regarding a minimum value.

\begin{figure}[h]
	\centering
        \includegraphics[width=\columnwidth]{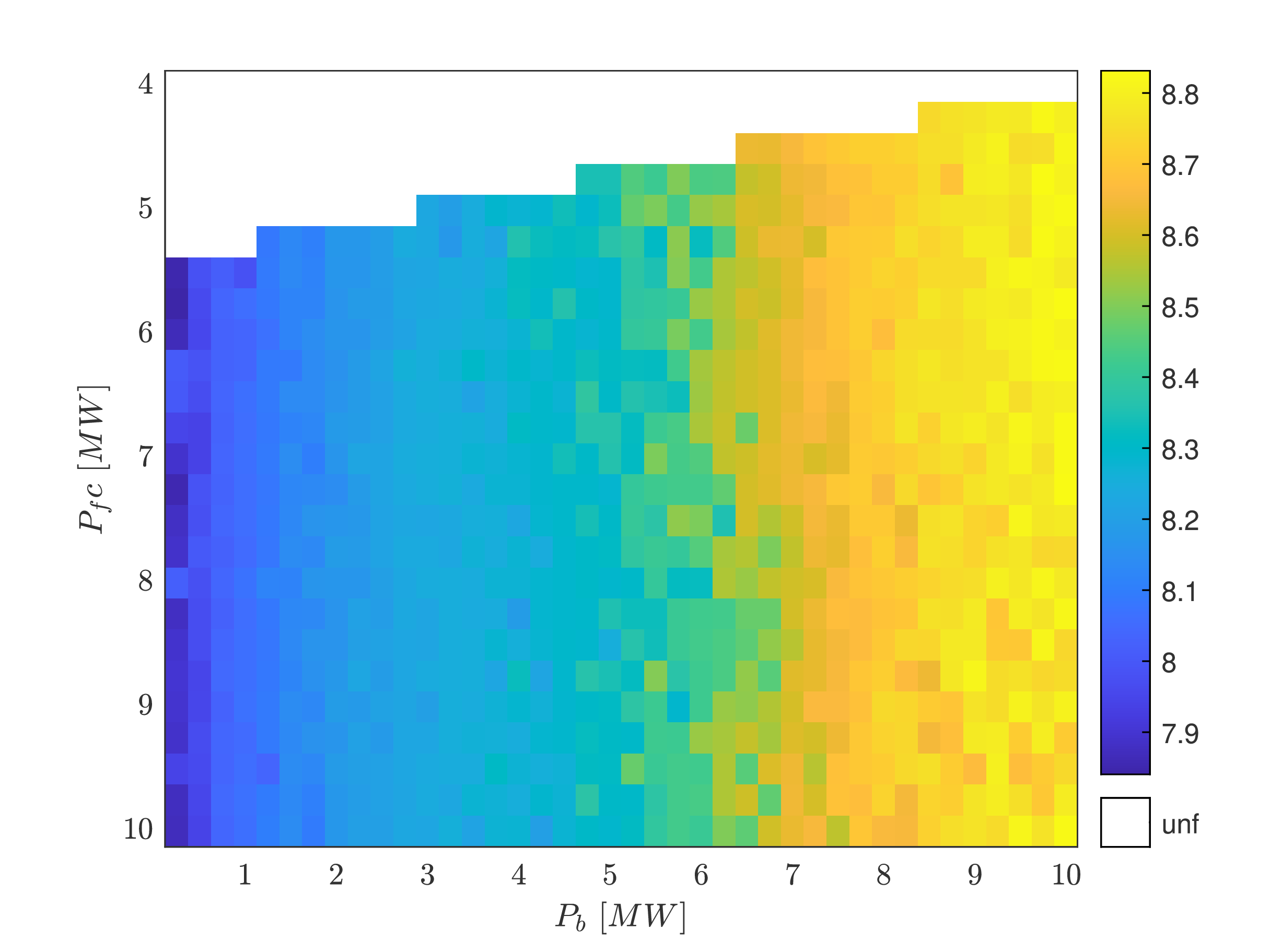}
	\caption{\ac{cii} comparison without CO\textsubscript{2} taxation.} \label{fig:CIIsensitivity_NO_TAX}
\end{figure}

\begin{figure}[h]
	\centering
        \includegraphics[width=\columnwidth]{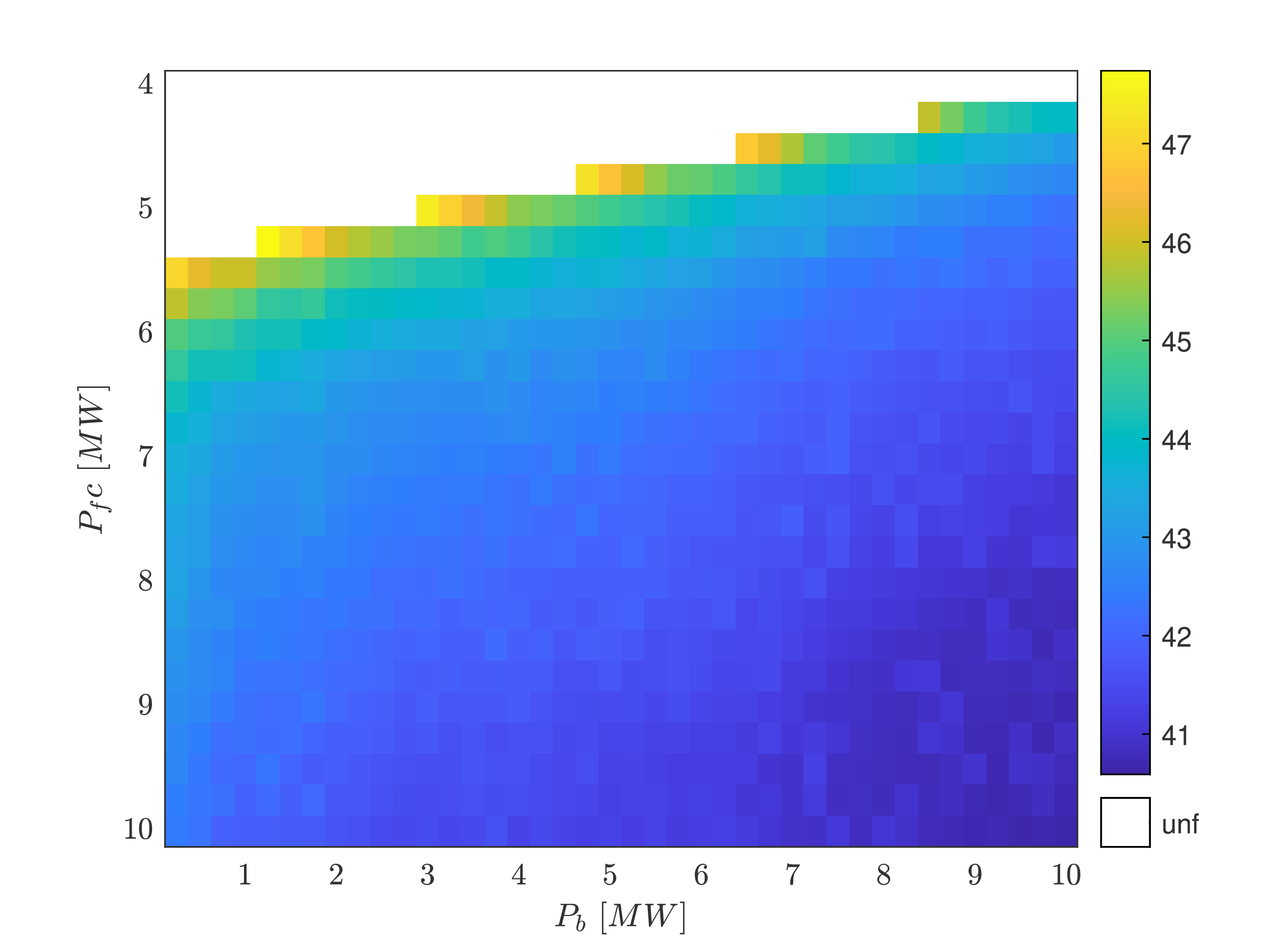}
	\caption{Cost comparison without CO\textsubscript{2} taxation.} \label{fig:COSTsensitivity_NO_TAX}
\end{figure}

\begin{figure}[h]
	\centering
        \includegraphics[width=\columnwidth]{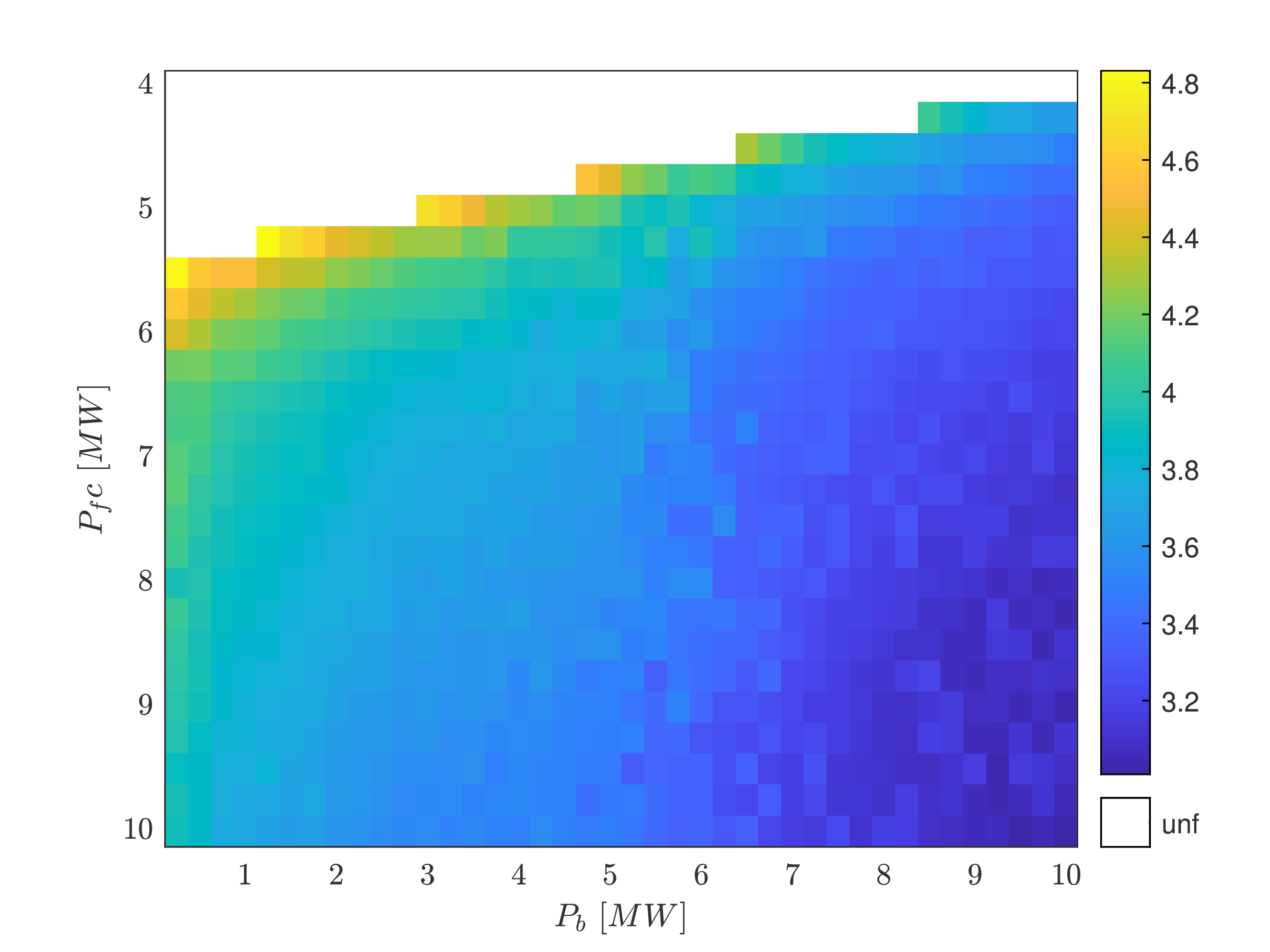}
	\caption{Hydrogen consumption comparison without CO\textsubscript{2} taxation.} \label{fig:FuelFCsensitivity_NO_TAX}
\end{figure}

To address the influence of the CO\textsubscript{2} taxation a comparison has been conducted. Figure \ref{fig:CIIsensitivity_NO_TAX} - \ref{fig:FuelFCsensitivity_NO_TAX} reports the results in which $c_{CO_2}$ is equal to zero.

\begin{figure}[h]
	\centering
        \includegraphics[width=\columnwidth]{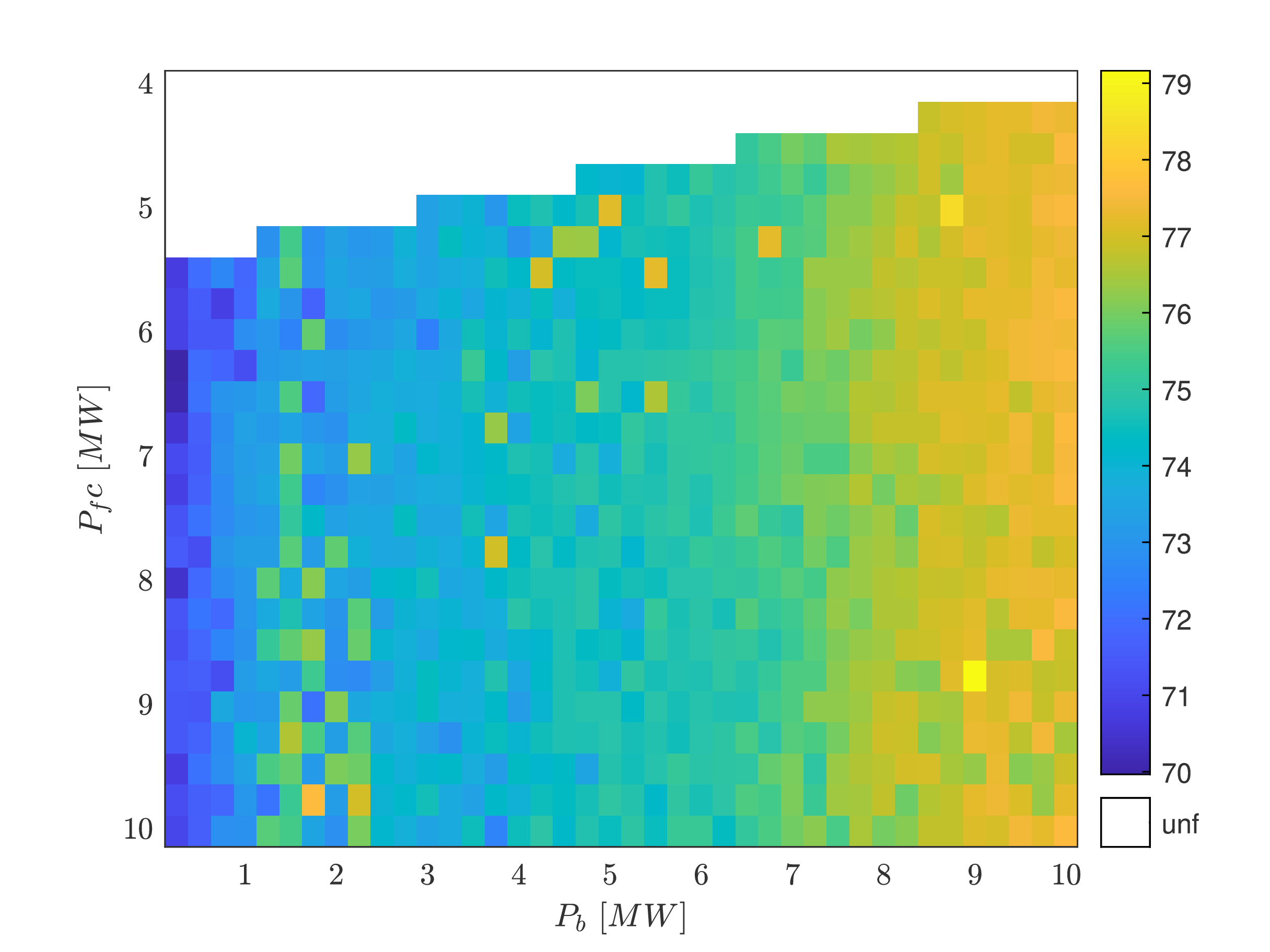}
	\caption{\ac{dg} average load factor comparison without CO\textsubscript{2} taxation.} \label{fig:LFsensitivity_NO_TAX}
\end{figure}

Overall, different pattern are presented from the previous analysis. In particular, it is possible to see that for battery energy ratings above \SI{5}{\MWh} the \ac{cii} becomes higher. In the other hand, to obtain an average load factor close to 80\% it is necessary to use a \ac{bess} with higher value of energy. 

\subsection{Study Cases}
The proposed power management strategy has been tested in two study cases. In the first study case (SC1), neither the \ac{bess} or the \ac{pemfc} are considered. In the second study case (SC2), both the \ac{bess} and the \ac{pemfc} are actively involved. It's worth noting that the Norwegian Maritime Authority mandates zero-emission navigation in the fjord. Therefore, in the second study case, the \ac{pms} is obligated to utilize only the battery and the fuel cell for the entire duration within the fjord. An exception is made for manoeuvring, where security constraints ensure that power generation is met by the \acp{dg} and the battery.

\begin{figure}[h]
	\centering
        \adjustbox{trim=0cm 0.6cm 0cm 0.2cm}{
        \includegraphics[width=\columnwidth]{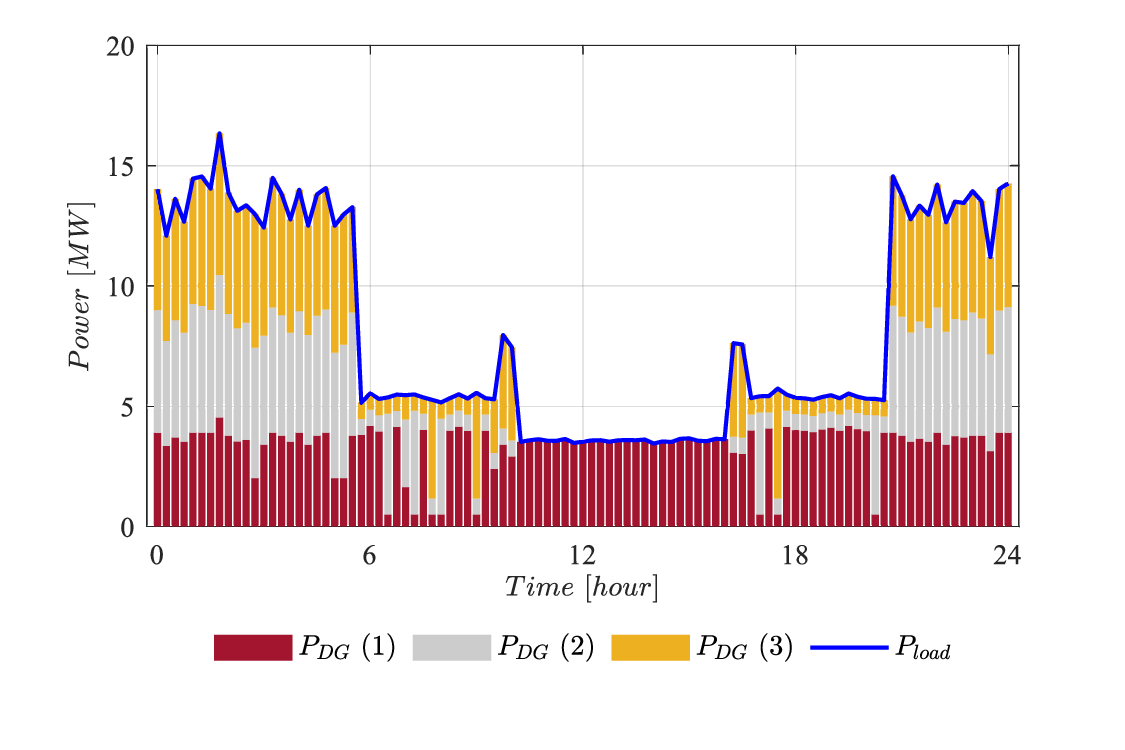}
        }
	\caption{Simulation results, SC1. Only \acp{dg}.} \label{fig:Fjord_DG}
\end{figure}

\begin{figure}[h]
	\centering
        \adjustbox{trim=0cm 0.6cm 0cm 0.2cm}{
        \includegraphics[width=\columnwidth]{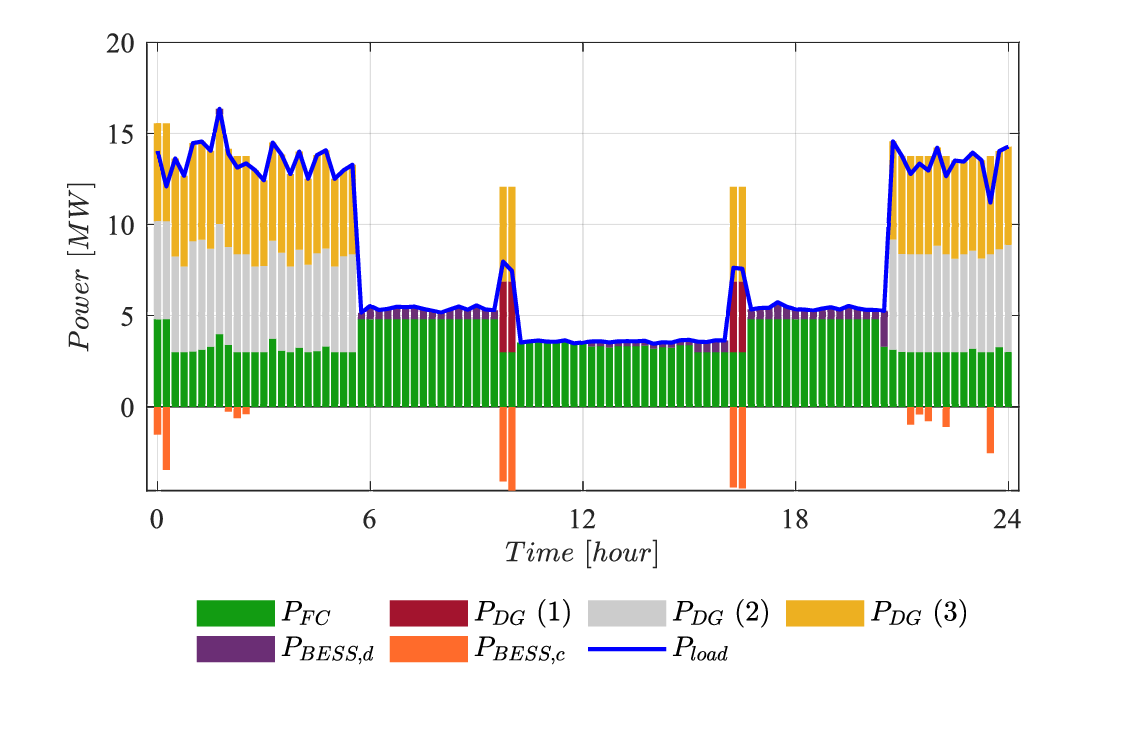}
        }
	\caption{Simulation results, SC2. \acp{dg}, \ac{pemfc} and \ac{bess}.} \label{fig:Fjord_Hybrid}
\end{figure}

Figure~\ref{fig:Fjord_DG} shows the results of  the optimization for the SC1, wherein the power generation is provided only by the \ac{dg}s. 
On the other hand, Fig. \ref{fig:Fjord_Hybrid} illustrates the results obtained for SC2, where both the \ac{bess} and the \ac{pemfc} are available.
The total power load, indicated in blue as $P_{load}$, represents the optimized load.


The results are summarized in Table \ref{tab:Results}, which reports the total cost, the amount of fuel and hydrogen required, the total CO\textsubscript{2} emission, the final value of the \ac{cii} and the average loading factor of the \acp{dg} ($LF_{avg}$).

\begin{table}[h]
    \centering
    \caption{Simulations results.}
    \label{tab:Results}
    {
    \begin{tabular}{cccccc}
        \midrule
        \midrule
        \multicolumn{6}{c}{SC1 - only \acp{dg}}\\
        \midrule
        Total Cost            & Total Fuel      & Total H\textsubscript{2}      & CO\textsubscript{2}   & CII                         & LF\textsubscript{avg}\\
        $[\SI{}{\EUR}]$       & $[kg]$          & $[kg]$                        & $[kg]$                & [$g/t \cdot nm$]            & [\%]\\
        \midrule
        65508                 & 35280           & 0                             & 113110                & 11.9                        & 58\\
        \midrule
        \midrule
        \multicolumn{6}{c}{SC2 - \acp{dg}, \ac{pemfc} and \ac{bess}}\\
        \midrule
        Total Cost            & Total Fuel      & Total H\textsubscript{2}      & CO\textsubscript{2}   & CII                         & LF\textsubscript{avg}\\
        $[\SI{}{\EUR}]$       & $[kg]$          & $[kg]$                        & $[kg]$                & [$g/t \cdot nm$]            & [\%]\\
        \midrule
        65575                 & 19029           & 5619                          & 61009                 & 6.4                         & 78.34\\
        \midrule
        \midrule
    \end{tabular}
    }
\end{table}

The cost of hydrogen considered is equal to \SI[per-mode=symbol]{154.29}{\EUR\per kWh}. This value includes the cost of fuel plus the compressed hydrogen storage costs reported in \cite{H2price}. Considering that the lower heating value of the hydrogen is equal to \SI[per-mode=symbol]{33.33}{kWh \per\kg}, $c_{H_2}$ is equal to \SI[per-mode=symbol]{5176}{\EUR\per\kg}, as previously reported (see Table \ref{tab:SimParam}).
The required \ac{cii} for the notional cruise ship is computed according to the \ac{imo} MEPC.336(76) standard \cite{IPD} and it is equal to 15 \SI{}{gCO_2/(t \cdot nm)}.
\begin{figure}[h]
	\centering
        \includegraphics[width=\columnwidth]{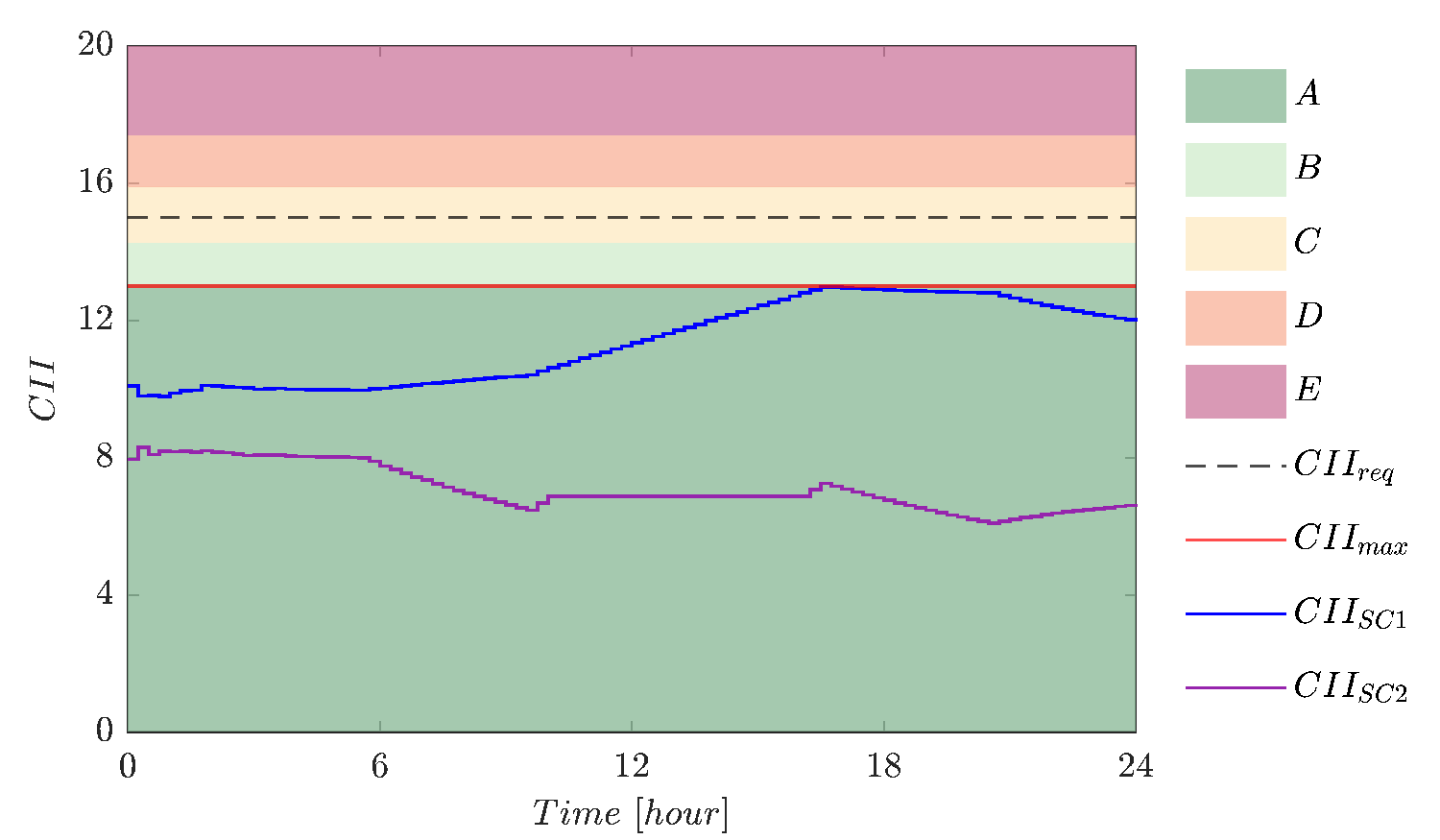}
	\caption{SC1 and SC2 \ac{cii} profile comparison.} \label{fig:CIIcomparison}
\end{figure}

The comparison of these two study cases shows that the \ac{bess} and the \ac{pemfc} allows to reduce the CO\textsubscript{2} emission by 46\% (\SI{52101}{\kg} of CO\textsubscript{2} saving with respect the SC1). 
In SC2, the average load factor ($LF_{avg}$) is in proximity to the minimum specific consumption, which occurs at around 80\% of the nominal power. It's worth noting that both case studies utilize the optimal control strategy.
The total cost is higher in SC2. This is due to the presence of the zero-emission constraint during navigation in the Fjord. In fact, hydrogen is more expensive than \ac{vlsfo}. The ranking levels are referred to the \ac{imo} MEPC.339(76) standard \cite{IPD}.

Figure \ref{fig:CIIcomparison} presents the comparison between the \ac{cii} profile in the two study cases. In SC1, the \ac{cii} reaches the maximum value imposed by the constraint, while in SC2, the \ac{cii} is significantly lower.

\begin{figure}[h]
	\centering
        \includegraphics[width=\columnwidth]{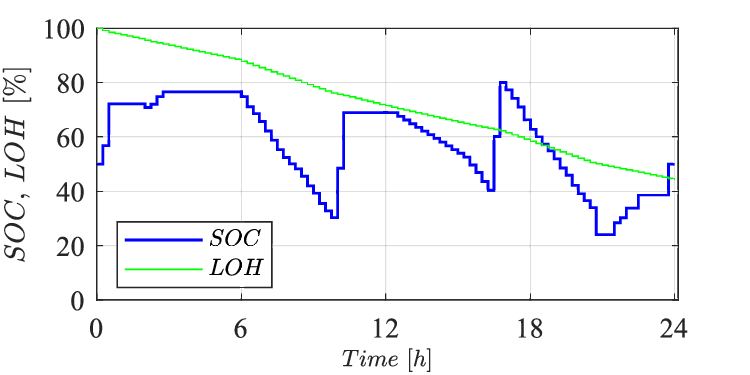}
	\caption{\ac{soc} and \ac{loh} comparison.} \label{fig:SOCandLOH}
\end{figure}

Figure \ref{fig:SOCandLOH} shows the comparison between the \ac{soc} of the battery and \ac{loh} of the hydrogen system. It is worth noting that the \ac{soc} is within its maximum (80\%) and minimum (20\%) limits.

\section{Conclusions}
This work proposes an optimal power management strategy for shipboard microgrids equipped with \acp{dg}, \ac{pemfc}s and \ac{bess}s.
The power management algorithm is based on a \ac{milp} problem that includes the \ac{uc} and the economic dispatch, to ensure a reliable power supply at minimum cost. A set of security constraints ensures to meet the \ac{iacs} requirements.

Two study cases have been investigated: the first one has been used as a reference as it considers only the utilization of \acp{dg}, while the second one includes also the exploitation of a \ac{pemfc} and a \ac{bess}. The available resources considered in the second study case allow reducing the CO\textsubscript{2} emission by 46\%, and enable zero-emission for specified operating conditions.

The sensitivity analysis highlights the importance of high power ratings for both \ac{fc} and \ac{bess} in achieving optimal zero emission capabilities, fuel efficiency, and operational cost savings. Lower hydrogen consumption is observed with specific power and energy ratings. This analysis offers valuable insights for optimizing system performance while considering feasibility constraints.

Future developments will be devoted to extend the algorithm through a model predictive control strategy.

\section*{Acknowledgment}
This research was partially funded by European Union – NextGenerationEU. Piano Nazionale di Ripresa e Resilienza, Missione 4 Componente 2 Investimento 1.4 “Potenziamento strutture di ricerca e creazione di ”campioni nazionali di R \& S” su alcune Key Enabling Technologies”. Code CN00000023 – Title: “Sustainable Mobility Center (Centro Nazionale per la Mobilità Sostenibile – CNMS)”. Views and opinions expressed are, however, those of the author(s) only and do not necessarily reflect those of the European Union or European Commission. Neither the European Union nor the granting authority can be held responsible for them.

\bibliographystyle{IEEEtran}
\bibliography{biblio}

\end{document}